\documentclass[10pt,letterpaper]{article}
\usepackage{opex3}
\usepackage{graphicx}
\begin{document}

\title{Quantum interferometry using coherent beam stimulated parametric down-conversion}

\author{Aziz Kolkiran and G S Agarwal}
\address{Department of Physics, Oklahoma State University,
Stillwater, OK - 74078, USA}
\email{aziz.kolkiran@okstate.edu}

\begin{abstract}
We show how stimulated parametric processes can be employed in
experiments on beyond the diffraction limit to overcome the
problem of low visibility obtained by using spontaneous down
conversion operating in the high gain regime. We further show
enhancement of the count rate by several orders when stimulated
parametric processes are used. Both the two photon counts and the
visibility can be controlled by the phase of the stimulating
coherent beam.
\end{abstract}



The question of beating the diffraction limit in optics has been
the subject of extensive discussions recently \cite{Boto
2000,Kapale 2007,Lee 2002,Agarwal 2001,Agarwal 2007,Liu
2007,Steinberg 2004,Walther 2004,O'Brien 2007,Kolkiran
2007,Kolkiran 2006,Steuernagel 2002,Leuchs 2007,Dowling
1998,Holland 1993,Ou 1997}. Dowling and coworkers proposed
\cite{Boto 2000} a very new idea to improve the sensitivity of
resolution by using detectors that work on two photon absorption
and by using special class of entangled states called $NOON$
states \cite{Kapale 2007} . They showed that the diffraction limit
can be beaten this way. The issue of the resolution in imaging
continues to be addressed \cite{Gatti 2003,
 Erkmen 2006, Bennik 2004, Dangelo 2001, Riberio 1996}.

It is easy to produce $NOON$ states experimentally with two
photons by using a very low gain parametric down converter. In
this case the resolution is improved by a factor of two. However
the probability of two photon absorption is very low unless one
could develop extremely efficient two photon absorbers. One
alternative would be to work with down converters in the high gain
limit \cite{De Martini 2008} however then the visibility of two
photon counts goes down asymptotically to $20\%$ \cite{Agarwal
2001}. Clearly we need to find methods that can overcome the
handicap of having to work with smaller visibility. Another
difficulty is with the magnitude of two photon counts. One needs
to improve the intensity of two photon counts considerably.

We propose a new idea using stimulated parametric processes along
with spontaneous ones \cite{Zeldovich 1969} to produce resolution
improvement while at the same time maintaining high visibility at
large gains of the parametric process. The stimulated processes
enhance the count rate by several orders of magnitude. We use
coherent beams at the signal and the idler frequencies. We further
find that the phases of coherent fields can also be used as tuning
knobs to control the visibility of the pattern. It may be borne in
mind that the process of spontaneous parametric down conversion
has been a work horse for the last two decades in understanding a
variety of issues in quantum physics and in applications in the
field of imaging \cite{Hong 1987,Shih 1994, Ou 1990, Rarity
1990,Kwiat 1995}.

\begin{figure}[h]
  \centering
  \scalebox{1.2}{\includegraphics{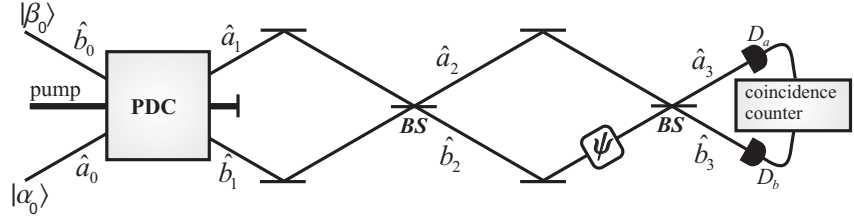}}
  \caption{Using an input from non-degenerate stimulated parametric down-conversion for determination of
  phase via photon-photon correlations.}
  \label{Fig1}
\end{figure}

We expect that the use of stimulated processes along with
spontaneous ones would change our landscape as far as fields of
imaging and quantum sensors are concerned. We now describe the
idea and the results of preliminary calculations that support the
above assertion. Consider the scheme shown in Fig. \ref{Fig1}.
Here $\hat{a}_1$ and $\hat{b}_1$ are the signal and idler modes
driven by the coherent fields. The usual case of spontaneous
parametric down conversion is recovered by setting
$\alpha_0=\beta_0=0$. The $\psi$ is the phase introduced by the
object or by an interferometer. For down conversion of type II the
signal and idler would be two photons in two different states of
polarization. In order to calculate the coincidence count it is
good to work with Heisenberg operators. The fields reaching the
detectors are related to the input vacuum modes $\hat{a}_0$ and
$\hat{b}_0$ via
\begin{eqnarray}\label{interferometer trans}
\left(\begin{array}{c}\hat{a}_3\\\hat{b}_3\end{array}\right)=\frac{1}{\sqrt{2}}\left(\begin{array}{cc}1
& i\\ i & 1
\end{array}\right)\left(\begin{array}{cc}1& 0\\ 0 &
e^{i\psi}\end{array}\right)
\frac{1}{\sqrt{2}}\left(\begin{array}{cc}1 & i\\ i & 1
\end{array}\right)\left(\begin{array}{c}\mu(
\hat{a}_0+\alpha_0)+\nu
(\hat{b}_0^{\dag}+\beta^*_0)\\\mu(\hat{b}_0+\beta_0)+\nu
(\hat{a}_0^{\dag}+\alpha^*_0)\end{array}\right),
\end{eqnarray}
 where $\mu$ and $\nu$ are given in terms of the gain parameter
$g$,
\begin{eqnarray}\label{bogolibov coeff}
\mu&=&\cosh(g),\\
\nu&=&e^{i\phi}\sinh(g).
\end{eqnarray}
and $\phi$ is the phase of the pump. We first note that in the
absence of the object $\psi=0$, the mean count say at the detector
$D_a$ is given by
\begin{equation}
I_{\hat{a}}\equiv\langle\hat{a}_3^{\dag}\hat{a}_3\rangle=\sinh^2(g)+
|\alpha_0|^2\left[1+2\sinh^2(g)+\sinh(2g)\cos(\phi-2\theta)\right],\label{single
counts a3}
\end{equation}
where for simplicity we assume that $\alpha_0=\beta_0$. We denote
$\theta$ as the phase of $\alpha_0$. Note that the first term in
Eq. (\ref{single counts a3}) is the intensity of spontaneously
produced photons. The $g$-independent term in the square bracket
is just the intensity of the coherent beam and the rest of the
terms result from stimulated parametric down conversion. Note
further that the mean count depends on the phase of the coherent
beams used to produce stimulated down conversion.

\begin{figure}[h]
  \centering
  \scalebox{.60}{\includegraphics{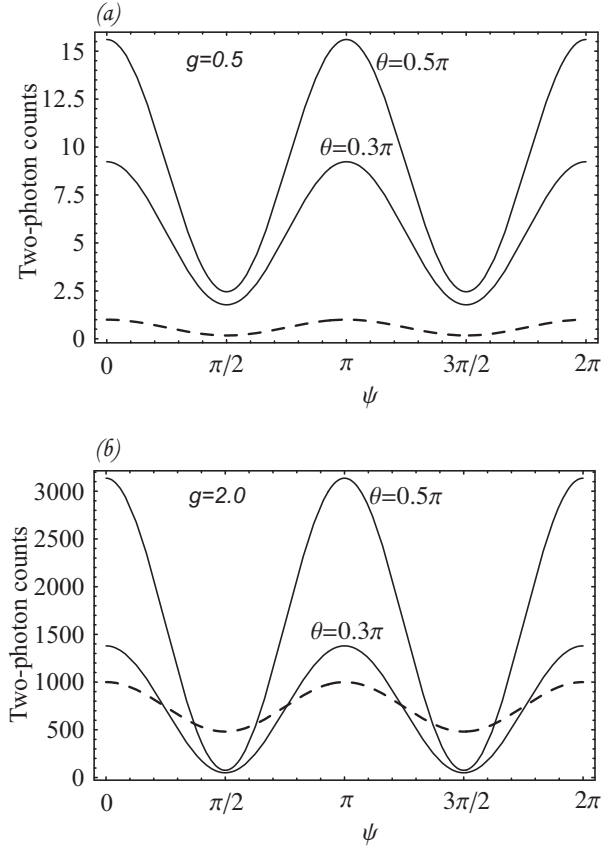}}
  \caption{($a$) Stimulated emission enhanced two-photon counts for various phases of the coherent field at the gain $g=0.5$.
  The horizontal line shows the interferometric phase. The pump phase  $\phi$ is fixed at $\pi$. The counts are in units of two-photon
   coincidence rates coming from spontaneous down-conversion process. The modulus of the coherent field $|\alpha|$ is chosen such that
   the coincidences coming from SPDC and the coherent fields are equal to each other. The dashed line shows the two-photon counts
    for the case of spontaneous process. ($b$) The same with (a) at the gain  $g=2.0$. Here, the counts for the case of spontaneous
    process (dashed line) is multiplied by a factor of $10^3$. }
  \label{Fig2}
\end{figure}

Now, using our basic equation (\ref{interferometer trans}) we
calculate the two-photon coincidence counts as the following:
\begin{equation}
I_{\hat{a}\hat{b}}\equiv\langle\hat{a}_3^{\dag}\hat{b}_3^{\dag}\hat{b}_3\hat{a}_3\rangle=A\biggl\{1+\frac{V}{1-V}\Bigl(1+\cos(2\psi)\Bigr)\biggr\}.
\label{two-photon counts}
\end{equation}
Here $V$ is the visibility of two-photon coincidence counts
\begin{equation}\label{visibility}
V=\frac{B}{A+B},
\end{equation}
where
\begin{eqnarray}\label{ABcoefficients}
    A&=&\sinh^4(g)+2|\alpha_0|^2\sinh^2(g)\left[1+2\sinh^2(g)+
\sinh(2g)\cos(\phi-2\theta)\right],\\
    B&=&\frac{1}{2}\biggl\{\left(1+\sinh^2(g)\right)\sinh^2(g)+|\alpha_0|^2\sinh(2g)\left[\sinh(2g)+\left(1+2\sinh^2(g)\right)\cos(\phi-2\theta)\right]\biggr.\nonumber\\
&+&\biggl.|\alpha_0|^4\left[1+2\sinh^2(g)+
\sinh(2g)\cos(\phi-2\theta)\right]^2\biggr\}.
\end{eqnarray}
Both $A$ and $B$ depend on the gain $g$, amplitude and phase of
the stimulating beams. In Figs. \ref{Fig2}$(a)$ and
\ref{Fig2}$(b)$ we display the fringes in two-photon counts under
different conditions on the gain of the down-converter and the
strength and phase of the stimulating beams.
These figures clearly show the advantages of using stimulating
parametric processes in quantum imaging. We next quantify these
advantages.

\begin{figure}[h]
  \centering
  \scalebox{.75}{\includegraphics{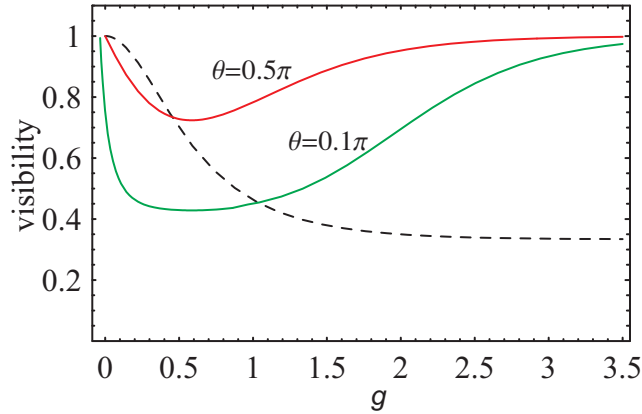}}
  \caption{(Color online) Stimulated emission enhanced visibility of two-photon
  counts for various
   phases (red and green lines) of the coherent field with respect to the gain $g$. The pump phase $\phi$  is fixed at $\pi$.
   The modulus of the coherent field $|\alpha_0|$ is chosen such that the coincidences coming from SPDC and
   the coherent fields are equal to each other. The dashed line shows the visibility of two-photon counts in the case of
   photons produced by spontaneous parametric down-conversion.}
  \label{Fig3}
\end{figure}

We first note that in the absence of stimulating fields
($|\alpha_0|\rightarrow 0$)
\begin{equation}
V\longrightarrow\frac{1+\sinh^2(g)}{1+3\sinh^2(g)},\label{SPDC
vis}
\end{equation}
and the strength of the two-photon counts reduces to
\begin{equation}
I_{\hat{a}\hat{b}}\longrightarrow2\sinh^4(g)+\sinh^2(g).\label{SPDC
strength}
\end{equation}
In the limit of large gain, the visibility drops to $1/3$ and the
strength of two-photon counts goes as $\exp(4g)$. Next, we examine
the effect of stimulated parametric processes on the visibility
and the numerical strength of two-photon coincidence count. In the
limit of large gain, the visibility of the stimulated process
reads
\begin{eqnarray}\label{sti visibility}
V\longrightarrow
\frac{\frac{1}{4}+|\alpha_0|^2\left(1+\cos(\Delta)\right)+|\alpha_0|^4\left(1+\cos(\Delta)\right)^2}
{\frac{3}{4}+3|\alpha_0|^2\left(1+\cos(\Delta)\right)+|\alpha_0|^4\left(1+\cos(\Delta)\right)^2},\nonumber\\
\end{eqnarray}
where $\Delta$ is the phase difference, $\phi-2\theta$, between
the pump and stimulating (coherent) beams. Note that when
$|\alpha_0|\rightarrow 0$ we recover the same result as Eq.
(\ref{SPDC vis}). The visibility given in Eq. (\ref{sti
visibility}) has  terms that arise from the interference between
the spontaneous and the stimulated down-converted photons. Clearly
we can control the value of the visibility by changing the
amplitude of the stimulating beams. For example, we can obtain
$60\%$ visibility even for $|\alpha_0|^2\sim1$ if $\Delta=0$,
which should be compared with the 33\% value in the absence of the
stimulating beams. As we increase the stimulating beam intensity
to $\sim10$, we obtain $90\%$ visibility. If we assume that the
stimulating field's intensity of the order of the number of
spontaneous photons produced by the down-converter, i.e.
$|\alpha_0|^2\sim\sinh^2(g)$, then the visibility of $100\%$ can
be reached at $g\simeq2-2.5$ (For $\Delta=\pi$ we lose the
advantage of stimulating beam to produce higher visibility.). In
Fig. \ref{Fig3}
\begin{figure}[h]
  \centering
  \scalebox{.75}{\includegraphics{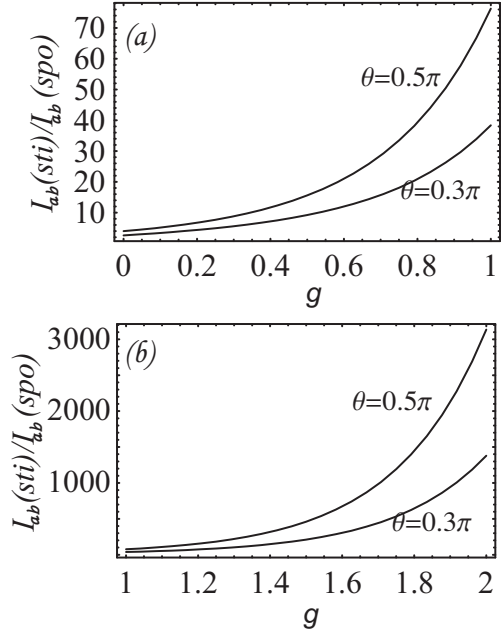}}
  \caption{The ratio of the two-photon coincidences coming from the stimulated process to the spontaneous
  process for various phases of the coherent beams
  at the ($a$) low and ($b$) high gain limits respectively. The pump phase is fixed at $\pi$ and
  the modulus of the coherent field $|\alpha|$ is chosen such that the coincidences coming from SPDC and
   the coherent fields are equal to each other.}
  \label{Fig4}
\end{figure}
we show the visibility of two-photon coincidences with respect to
the gain for different values of the stimulating beam phases. The
results in the region of large gain follow the approximate results
based on Eq. (\ref{sti visibility}).

We next examine the strength of two-photon counts in the limit of
high gain. This depends on the interferometric phase $\psi$. To
get an estimate of the strength of two-photon counts let us set
$\psi=0$:
\begin{equation}\label{sti strength}
I_{\hat{a}\hat{b}}\longrightarrow\,\,2\sinh^4(g)\biggl\{
1+4|\alpha_0|^2\left(1+\cos(\Delta)\right)\biggr.
 \biggl.+2|\alpha_0|^4\left(1+\cos(\Delta)\right)^2 \biggr\}.
\end{equation}
Note that when $\alpha_0=0$ we recover Eq. (\ref{SPDC strength}).
 For $\Delta=0$, the highest order term in Eq. (\ref{sti strength})
goes as $\exp(4g)|\alpha_0|^4$, i.e. a factor of $|\alpha_0|^4$
appears here in compared to the spontaneous process. This then
reduces to $I_{\hat{a}\hat{b}}\rightarrow\exp(8g)$ if we assume
that the stimulating field's intensity of the order of the number
of spontaneous photons produced by the down-converter, i.e.
$|\alpha_0|^2\sim\sinh^2(g)$. This leads to an enhancement by
$\exp(4g)$ in the two-photon count rates compared to the case of
spontaneous processes. In Figs. \ref{Fig4}$(a)$ and
\ref{Fig4}$(b)$, we show the ratio of two-photon counts coming
from the stimulated process to the spontaneous process both at the
low and high gain limits respectively. It is shown that at
$g\simeq 1.7$, three orders of magnitude rate enhancement is being
reached. Therefore, in the determination of interferometric phase,
we obtain a ground-breaking  enhancement in both the visibility
and the strength of the two-photon coincidence counts by
controlling the phase and the amplitude of stimulating coherent
beams. We show in Figs. \ref{Fig2}$(a)$ and \ref{Fig2}$(b)$, this
cumulative enhancement in both the visibility and the strength in
the low and high gain limits respectively.

A question that we have not investigated in the present paper
concerns the minimum value of the phase $\Delta\psi$ that can be
measured \cite{Leuchs 2007}. In the literature one has the well
known shot noise limit ($\Delta\psi\sim 1/\sqrt{N}$; where $N$ is
the total number of photons) obtained with coherent sources. This
is to be compared with the Heisenberg limit ($\Delta\psi\sim 1/N$)
obtained with sources prepared in special states and with very
special detection schemes \cite{Kok 2002, Gerry 2003}. Thus to
improve the sensitivity it would be especially interesting if one
can do the latter with photon numbers of the same order as in
coherent sources. However so far one has achieved Heisenberg limit
only with photon numbers of order few. Thus the real question
is--what is the achievable phase uncertainty given the presently
available sources and measurement techniques. This is something
that needs to be studied at depth. We note that the original
proposal of Dowling and collaborators employed the $NOON$ states
and measurements based on the observable $|N0\rangle\langle
0N|+|0N\rangle\langle N0|$. There have been other suggestions
which enable one to achieve Heisenberg limit. Some of these are
based on homodyne measurements \cite{Steuernagel 2004} whereas
others \cite{Smerzi 2008} make use of in principle measurements
which would achieve Cramer-Rao lower bound on phase sensitivity.
It would clearly be interesting to generalize the latter proposals
when stimulating signal and idler fields are employed.

In conclusion, we have shown that using stimulated parametric
processes along with spontaneous ones leads to resolution
improvement and high signal values while at the same time
maintaining high visibility at large gains of the parametric
process. We use coherent beams at the signal and idler
frequencies. We find that the phases of coherent fields can also
be used as tuning knobs to control the visibility of the pattern.
The use of stimulated parametric down-conversion  also improves
the rates of two-photon absorption in quantum lithography. The use
of stimulated processes in multi-photon coincidence events is
expected to produce even bigger advantages, for example in
producing much higher count rates. We hope to examine these in
future. Finally we believe that the use of stimulated processes
along with spontaneous ones would change our landscape as far as
fields of imaging and quantum sensors are concerned.



 \end{document}